\documentstyle[11pt,newpasp,twoside,epsf]{article} % dvips users ONLY

\def\dd{{\rm d}}

\def\be{\beta}
\def\Gn{\Gamma_n}
\def\bn{\beta_n}
\def\Gej{\Gamma_{\rm ej}}

\def\Gsh{\Gamma_{\rm sh}}

\def\Mej{M_{\rm ej}}

\def\Rdec{R_{\rm dec}}
\def\Rtr{R_{\rm trail}}

\def\mdec{m_{\rm dec}}
\def\be{\begin{equation}}
\def\ee{\end{equation}}

\def\dM{\dot{M}}

\def\Rb{R_\beta}
\def\mb{m_\beta}
\def\taub{\tau_\beta}

\def\dM{\dot{M}}

\newbox\grsign \setbox\grsign=\hbox{$>$} \newdimen\grdimen \grdimen=\ht\grsign
\newbox\simlessbox \newbox\simgreatbox \newbox\simpropbox
\setbox\simgreatbox=\hbox{\raise.5ex\hbox{$>$}\llap
     {\lower.5ex\hbox{$\sim$}}}\ht1=\grdimen\dp1=0pt
\setbox\simlessbox=\hbox{\raise.5ex\hbox{$<$}\llap
     {\lower.5ex\hbox{$\sim$}}}\ht2=\grdimen\dp2=0pt
\setbox\simpropbox=\hbox{\raise.5ex\hbox{$\propto$}\llap
     {\lower.5ex\hbox{$\sim$}}}\ht2=\grdimen\dp2=0pt
\def\simgt{\mathrel{\copy\simgreatbox}}

\def\edcomment#1{\iffalse\marginpar{\raggedright\sl#1\/}\else\relax\fi}
\reversemarginpar

\begin{document}
\title{Fireballs with a Neutron Component}
\author{Andrei M. Beloborodov}
\affil{Canadian Institute for Theoretical Astrophysics, 
University of Toronto, 60 St. George Street, Toronto, Ontario M5S 3H8,
Canada}

\begin{abstract}
Standard GRB fireballs must carry free neutrons. This crucially 
changes the mechanism of fireball deceleration by an external medium. As 
the ion fireball decelerates, the coasting neutrons form a leading front. 
They gradually decay, leaving behind a relativistic trail of decay products 
mixed with the ambient medium. The ion fireball sweeps up the trail and 
drives a shock wave in it. Thus, observed afterglow emission is produced 
in the neutron trail. The impact of neutrons turns off at $\sim 10^{17}$~cm 
from the explosion center, and here a spectral transition is expected in 
GRB afterglows. Absence of neutron signatures would point to absence of 
baryons and a dominant Poynting flux in the fireballs.
%in GRB ejecta and strongly magnetized fireballs.
%in the fireballs and a strong dominance of a Poynting flux.
\end{abstract}

%#######################################################################

\section{Introduction}

Importance of neutrons in GRBs was realized recently
(Derishev, Kocharovsky, \& Kocharovsky 1999a,b; Bahcall \& M\'esz\'aros 
2000; M\'esz\'aros \& Rees 2000; Fuller, Pruet, \& Abazajian 2000; 
Pruet \& Dalal 2002; Beloborodov 2003a,b). 
Their presence is inevitable in a standard fireball (sections 2 and 3) 
and they qualitatively change the explosion picture at radii of
$10^{16}-10^{17}$~cm (section~4).

%#######################################################################

\section{Neutronization of the Central Engine}

GRB central engines are very dense and hot (e.g. M\'esz\'aros 2002). 
The central matter (or matter surrounding the central black hole) is 
filled with hot $e^\pm$ pairs in thermodynamic equilibrium with blackbody 
radiation at a temperature $kT=1-10$~MeV. Its baryonic component is made
of neutrons and protons (nuclei break up into free nucleons in the 
unshadowed region of the $T-\rho$ plane shown in left panel of Fig.~1) 
and frequent $e^\pm$ captures on nucleons take place,
\begin{equation}
  e^-+p\rightarrow n+\nu, \qquad e^++n\rightarrow p+\bar{\nu}.
\end{equation}
%(Neutrinos are emitted in these reactions, which is the dominant cooling
%mechanism of the GRB engines.) 
These reactions quickly convert protons to neutrons and back, and establish 
an equilibrium proton fraction $Y_e=n_p/(n_n+n_p)$ which is shown in Figure~1. 
%Neutrons dominate ($Y_e<0.5$) and if $T<T_n=W\rho_{11}^{1/2}$~MeV, where
%$W=33$~MeV for $\nu$-opaque matter and $W=21$~MeV for $\nu$-opaque matter.
Any plausible GRB engines belong to the $T-\rho$ region where $Y_e<0.5$.
%For $\nu$-opaque matter, reactions 
%inverse to (1) should be included, which slightly changes this condition
%(B03a). At any rate, plausible GRB engines satisfy the condition. 
So, the neutrons are not only present in the central engines --- in fact, 
they dominate.
% over the protons.

When the neutronized matter is ejected into a high-entropy fireball, it 
remains neutron-dominated because (1) expansion occurs too fast 
and $Y_e$ freezes out at its initial value below 0.5, or (2) the fireball 
partially absorbs the neutrino flux from the engine, which tends to keep 
$Y_e<0.5$ (see Beloborodov 2003a for details).
%because $\bar{\nu}$ are likely more energetic than $\nu$ and fi

%%%%%%%%%%%%%%%%%%%%%%%%%%%%%%%%%%%%%%%%%%%%%%
\begin{figure}
%\plotone{beloborodova_1.eps}
\plottwo{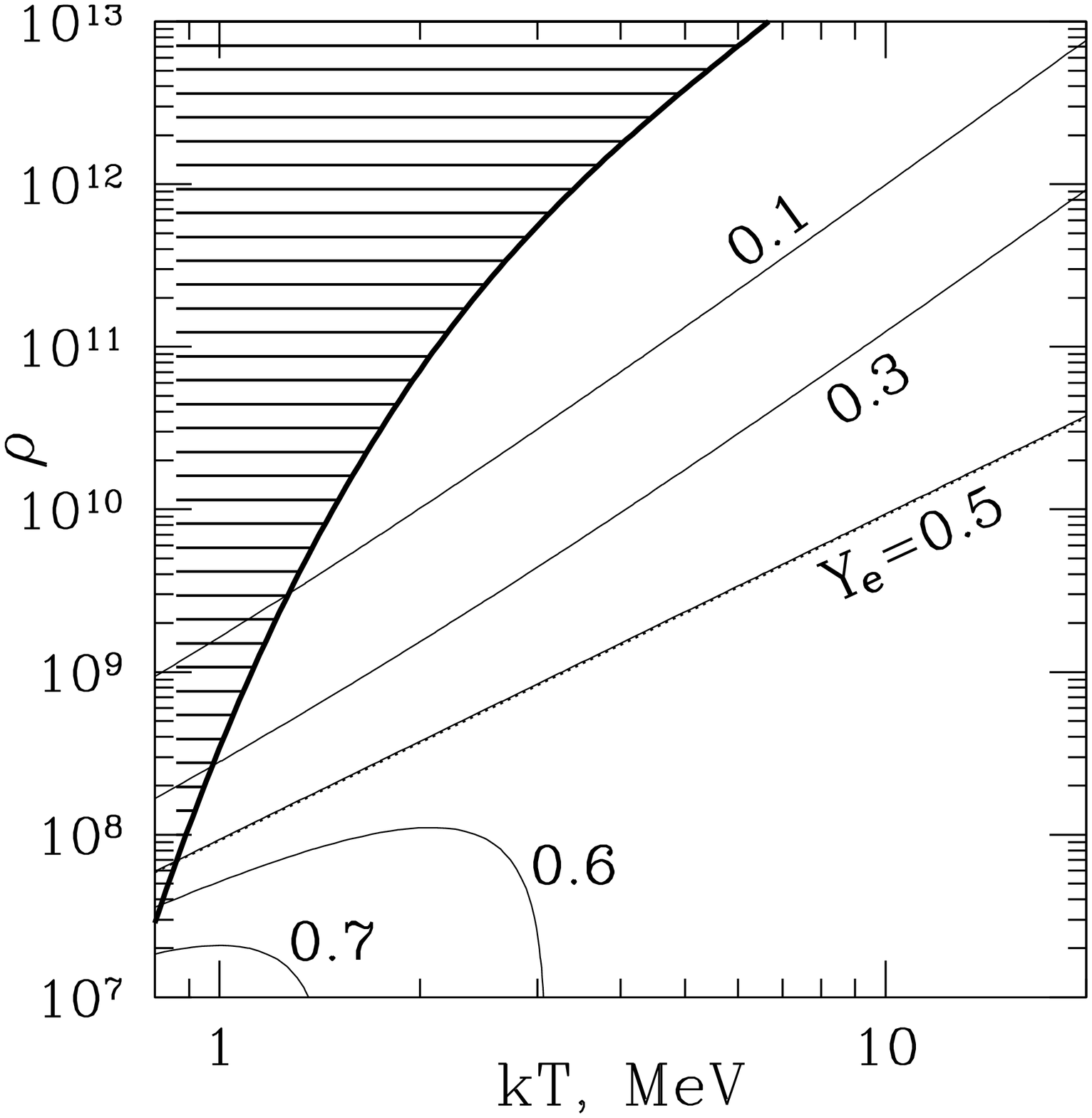}{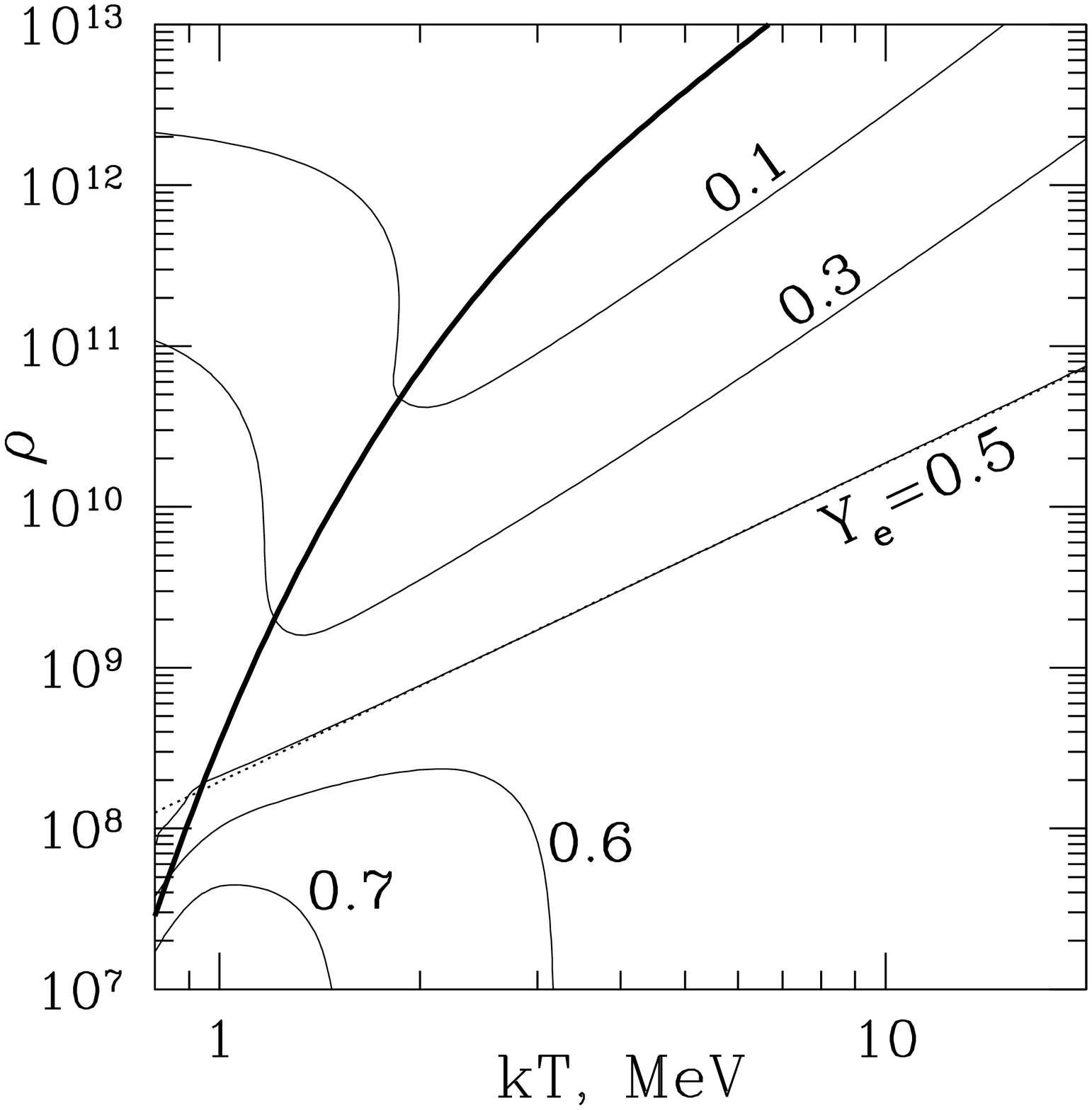}
%\epsf{figure=beloborodova1_1.eps,width=0.5\textwidth}
%\psfig{figure=fig1b.ps,width=0.5\textwidth}
\caption{Contours of equilibrium $Y_e=n_p/(n_n+n_p)$ on
the $T$-$\rho$ plane for $\nu$-transparent (left) and $\nu$-opaque (right)
matter. Thick curve shows the boundary of free-nucleon region.
(From Beloborodov 2003a)
}
\end{figure}
%%%%%%%%%%%%%%%%%%%%%%%%%%%%%%%%%%%%%%%%%%%%%%

%#######################################################################

\section{Ejected Neutrons Survive the Nucleosynthesis}

As the fireball expands and cools, nucleons tend to
recombine into $\alpha$-particles (like they do in the big bang). 
This process competes, however, with rapid expansion and can freeze out. 
For this reason, nucleosynthesis is suppressed in fireballs with a high 
photon-to-baryon ratio $\phi=n_\gamma/n_b$ (or, equivalently, high entropy 
per baryon $s/k=3.6\phi$). A minimum $\phi$ in GRBs is $\sim 10^5$, which is
just marginal for nucleosynthesis (detailed calculations of nucleosynthesis 
in GRBs have been done recently by Lemoine 2002; Pruet, Guiles, \& Fuller 
2002; Beloborodov 2003a).
%We find that, in radial fireballs, more than half of nucleons can recombine 
%only if $\phi<3\times 10^4(r_0/3\times 10^6)^{0.85}$ where $r_0$ [cm] is 
%the size of the central engine. In fireballs with parabolic collimation, 
%the efficient recombination requires
%$\phi<6\times 10^5(r_0/3\times 10^6)^{0.9}$. 
Even in the extreme case of complete nucleon recombination there 
are still leftover neutrons because of the neutron excess ($Y_e<0.5$). 
Their minimum mass fraction is $X_n=1-2Y_e$.

In addition, synthesized helium is likely destroyed during the subsequent 
evolution of the explosion. This happens if (1) there appears a substantial 
relative bulk velocity between the neutron and ion components (as a result 
of neutron decoupling during the acceleration stage of the fireball) or 
(2) internal shocks occur and heat the ions to a high temperature 
(Beloborodov 2003a).

Thus, a substantial neutron component appears inevitably in the standard 
fireball scenario. Neutrons develop a Lorentz factor $\Gn=10^2-10^3$ at the 
very beginning of the explosion when the fireball is accelerated by 
radiation pressure: they are collisionally coupled to the ions in the
early dense fireball, and decouple close to the end of the acceleration 
stage. Then the neutrons coast and gradually decay with a mean lifetime
$\taub\approx 900$~s and a mean decay radius $\Rb=c\taub\Gn$,
\be
\label{eq:H}
 \Rb=0.8\times 10^{16}\left(\frac{\Gn}{300}\right) {\rm ~cm}.
\ee

%#######################################################################

\section{Neutron-Fed Blast Wave}

Let us remind what happens in a relativistic explosion without neutrons. 
The ejected fireball with mass $\Mej$ and Lorentz factor $\Gej$ sweeps up 
an ambient medium with density $n_0=1-100$~cm$^{-3}$ and gradually dissipates 
its kinetic energy. The dissipation rate peaks at a characteristic 
``deceleration'' radius 
$\Rdec\sim 10^{16}$~cm where half of the initial energy is dissipated.
$\Rdec$ corresponds to swept-up mass $\mdec=\Mej/\Gej$. Further
dynamics is described by the self-similar blast wave model of Blandford \&
McKee (1976). How does this picture change in the presence of neutrons?

At radii under consideration, $R>10^{15}$~cm, the ejected fireball is a
shell of thickness $\Delta\ll R$. In contrast to neutrons, the ion component
of the fireball is aware of the external medium and its Lorentz factor
$\Gamma$ decreases. As $\Gamma$ decreases below $\Gamma_n$, the ions fall
behind and separate from the neutrons. Thus the fireball splits into two
relativistic shells which we name N (neutrons) and I (ions).
%%; the N-shell will also be called ``neutron front''. 
% The I-shell lags behind by a distance $l\approx R(1/2\Gamma^2-1/2\Gn^2$,
% which much smaller than $R$ but soon becomes larger than $\Delta$.
%%\be
%%\label{eq:l}
% \frac{l}{R}\approx \bn-\bh \approx \frac{1}{2\Gamma^2}-\frac{1}{2\Gn^2},
%\ee
%where $\bh$ and $\bn$ are the shell velocities in units of $c$.
%For simplicity, we hereafter focus on the stage of complete separation
%$l>\Delta$ (it sets in right after the beginning of the I-shell deceleration
%if $\Delta\sim 10^{11}$~cm, and it covers the whole explosion if
%$\Delta\rightarrow 0$).
The mass of the leading N-shell is decreasing because of the $\beta$-decay,
\be
\label{eq:Mn}
  M_n(R)=M_n^0\exp\left(-\frac{R}{\Rb}\right).
\ee
The N-shell energy, $E_n=\Gn M_nc^2$, is huge compared to the ambient 
rest-mass $mc^2$ even at $R>\Rb$. For example, at $R=\Rdec$ we
find $E_n/\mdec c^2=X_n\Gn\Gej\exp(-\Rdec/\Rb)$ where $X_n=M_n^0/\Mej$ is 
the initial neutron fraction of the fireball.

The neutron decay products $p$ and $e^-$ share immediately their huge
momentum with ambient particles due to two-stream instability
%(the instability timescale is set by the ion plasma frequency
%$\omega_p$ and it is the shortest timescale in the problem). Thus, 
and the  N-shell leaves behind a mixed trail with a relativistic
bulk velocity $\beta<\bn$ (Beloborodov 2003b)
\be
\label{eq:beta}
  \beta(R)=\frac{\bn}{1+(\Gn\zeta)^{-1}},  \qquad
  \gamma(R)=\frac{1}{(1-\beta^2)^{1/2}}
           =\frac{\Gn\zeta+1}{(\zeta^2+2\Gn\zeta+1)^{1/2}},
\ee
where 
\be
\label{eq:zet}
  \zeta(R)=\frac{\dd M_n}{\dd m}
              =\frac{M_n}{\Rb}\left(\frac{\dd m}{\dd R}\right)^{-1},
\ee
and $m(R)$ is ambient mass enclosed by radius $R$. There exists
a characteristic radius $\Rtr$ where the trail becomes nonrelativistic
($\beta=0.5$). It is defined by condition $\zeta=\Gn^{-1}$,
which requires about 10 e-folds of the decay
(for a typical $\mb\sim\mdec\sim 10^{-5} M_n^0$). Thus,
\be
 \Rtr\approx 10\Rb=0.8\times 10^{17}\left(\frac{\Gn}{300}\right){\rm ~cm}.
\ee
$\Rtr$ depends very weakly (logarithmically) on the ambient density and the
initial neutron fraction of the fireball, $X_n$.

The decaying N-shell not only accelerates the medium as it passes through
it. It also compresses the medium, loads with new particles, and heat 
to a high temperature. The rest-frame density and relativistic 
enthalpy of the trail are
%Given an initial ambient density $n_0$ (ahead of the neutron front),
%the rest-frame density of the trail, $n$, is easily calculated 
\be
\label{eq:mu}
  n=n_0(1+\zeta)\left(\zeta^2+2\Gn\zeta+1\right)^{1/2}, \qquad
  \mu=\frac{(\zeta^2+2\Gn\zeta+1)^{1/2}}{1+\zeta},
\ee
where $n_0$ is the medium density ahead of the N-shell. 
For $\Gn^{-1}<\zeta<\Gn$ one finds $\mu\gg 1$, i.e. 
the thermal energy of the trail far exceeds its rest-mass energy.

The ion fireball follows the neutron front and collects the trail.
As a result, (1) the ion Lorentz factor $\Gamma$ decreases and (2) a shock
wave propagates in the trail material. The shock has a Lorentz factor 
$\Gsh\simgt\Gamma$ and it cannot catch up with the neutron 
front (unless $n_0[R]$ falls off steeper than $R^{-3}$). Dynamics and 
dissipation in the shock are discussed in Beloborodov (2003b). We emphasize 
here important differences from a customary external shock: the 
neutron-trail shock propagates in a relativistically moving, dense, hot, 
and possibly magnetized medium behind the leading neutron front. 
%The fate of magnetic fields in such shocks and their emission is 
%an interesting issue for a future study.

The neutron impact ceases at $\Rtr\approx 10^{17}$~cm, which can leave an
imprint on the observed afterglow. For example, the shock dissipation can
have a second bump (Beloborodov 2003b), and a spectral transition is also 
possible. The arrival time of radiation emitted at $\Rtr$ is approximately
$\Rtr/2\Gamma^2 c$ (counted from the arrival of first
$\gamma$-rays). It may be as long as 30 days or as short as a few seconds,
depending on the fireball Lorentz factor $\Gamma(\Rtr)$. 
%Remarkably, $\Rtr$ is
%almost independent of the ambient medium, and its observational signature
%would give information on the neutron Lorentz factor. 
Recent early observation of a GRB afterglow (GRB~021004) discovered an 
interesting re-brightening at $10^3$~s.
%Future systematic studies by {\it Swift} will
%show whether such features are common and allow useful tests. 
Also, we do not exclude a possible relevance of neutrons to the 20~day 
bumps observed in a few GRBs, as the time coincides with $\Rtr/c$.

Neutron signatures should be absent if the fireball is 
dominated by a Poynting flux and has extremely low baryon loading. Then the 
neutron component decouples early, with a modest Lorentz factor $\Gn$, and 
decays at small radii. The upper bound on $\Gn$ due to decoupling is 
$\Gn\approx 300(\dM_\Omega/10^{26})^{1/3}$ where $\dM_\Omega$ [g/s] is the 
mass outflow rate per unit solid angle of the fireball (Beloborodov 2003a).

An additional piece of GRB physics is the interaction of the prompt 
$\gamma$-ray radiation with an ambient medium (see Beloborodov 2002 and
refs. therein). It leads to a gap opening and crucially affects the 
early afterglow emitted at radii $R<10^{16}(E_\gamma/10^{53})^{1/2}$~cm, 
where $E_\gamma$ [erg] is the isotropic energy of the GRB. Here, we focused 
on larger radii where the neutron effects dominate.  

%A standard fireball model has   
%We focused here on the neutron front and did not account for the
%$\gamma$-ray precursor that impacts the blast wave dynamics at
%$R<\Racc=0.7\times 10^{16}(E_\gamma/10^{53})^{1/2}$~cm, where $E_\gamma$
%[erg] is the isotropic energy of the GRB (see Beloborodov 2002 and
%refs. therein). The analysis in this Letter is strictly valid for
%afterglows emitted at $R>\Racc$. Then the radiation-front effects,
%including the gap opening at $R<0.3\Racc$, occur at smaller radii, and
%apply to the earlier afterglow. For a dense medium, where $\Rdec<\Racc$,
%effects of the neutron and $\gamma$-ray fronts should be studied together.

%\section{Conclusions}
%
%Besides making the fireball an interesting source of multi-GeV
%neutrinos (Derishev et al. 1999, Bahcall \& M\'esz\'aros 2000, M\'esz\'aros
%\& Rees 2000) the neutrons survive and play a dramatic role for the
%explosion development at large radii $R\sim 10^{16}-10^{17}$~cm as
%explained below.

%\acknowledgments

%#######################################################################

%Journal names and book titles should be set in {\it italics\/}.
%Volume numbers should be {\bf boldface}.

\end{document}